# Anomalous light absorption around subwavelength apertures in metal films

O. Lozan[1], M. Perrin[1], B. Ea-Kim[2], J.M. Rampnoux[1], S. Dilhaire[1], and P. Lalanne[3]

[1]Laboratoire Onde et Matière d'Aquitaine (LOMA) UMR 5798, CNRS-Université de Bordeaux, 33400 Talence, France

[2]Laboratoire Charles Fabry, UMR 5298, CNRS-IOGS-Université Paris XI, Institut d'Optique, 91120 Palaiseau, France.

[3]Laboratoire Photonique, Numérique et Nanosciences (LP2N), UMR 5298, CNRS-IOGS-Université de Bordeaux, Institut d'Optique d'Aquitaine, 33400 Talence, France.

PACS numbers:

42.25.Fx Diffraction and scattering
78.68.+m optical properties of surfaces
73.20.Mf Collective excitations
78.47.D- Time resolved spectroscopy

**Abstract**

In this letter, we study the heat dissipated at metal surfaces by the electromagnetic field scattered by isolated subwavelength apertures in metal screens. In contrast to our common belief that the intensity of waves created by local sources should decrease with the distance from the sources, we reveal that the dissipated heat at the surface remains constant over a broad spatial interval of a few tens of wavelengths. This behavior which occurs for noble metals at near infrared wavelengths is observed with nonintrusive thermoreflectance measurements and is analyzed with an analytical model which underlines the intricate role played by quasi-cylindrical waves in the phenomenon. Additionally, computational results reveal that the absorption plateau is an intrinsic property that occurs for a large variety of topologies, irrespective of aperture shapes.

Dynamical diffraction teaches us that the intensity of waves created by local sources in uniform and lossless media gradually decreases with the distance from the sources [1]. This property is valid for all waves that carry energy, from exotic gravitational waves to all sounds of our daily life. In the vicinity of surfaces, the near-field propagation is altered by additional boundary conditions, nevertheless one again expects the wave intensity on the surface to gradually decrease with the distance. For instance, the electromagnetic or acoustic intensities scattered by subwalength slits (or holes) at the surface of perfect electromagnetic conductors or incompressible materials decreases algebraically as $1/x$ (or $1/x^2$) with the distance $x$ from the aperture. For finite electrical conductivity, absorption additionally comes into play, reinforcing our expectation for a monotonous intensity damping.

In this letter we report on a remarkably simple case that invalidates this intuition. This occurs for the field scattered by isolated subwavelength apertures on the surface of noble metal films at near infrared frequencies ($\lambda > 0.8$). Transmission and scattering by such apertures in metal films is a fundamental phenomenon of wave physics with a long history [2]. In recent years, the topic has been a subject of renewed interest because of its importance in modern sciences and techniques [3-6]. However, despite these efforts, little is presently known about the heat dissipated by the electromagnetic field in the metal surface around the aperture. Actually, we show that, as it propagates away from the slit, this field heats the metal and paradoxically the heat dissipated at the surface remains constant over a broad spatial scale of several tens of wavelengths. The phenomenon is general since the *same* absorption plateau occurs for a variety of topologies, irrespective of whether the aperture shape corresponds to one-dimensional line (slit) or point-like (hole) sources. It is evidenced by thermoreflectance measurements and well predicted by a simple analytical model based on a recent work that describe the field radiated on metal surfaces as a superposition of two waves, a surface-plasmon polaritons and a quasi-cylindrical wave [7]. The model additionally shows that the heat dissipated in the plateau area is actually weaker than the heat deposited by the plasmon launched on the surface. This highlights the intricate role played by the quasi-cylindrical wave at the metal surface.

For the thermoreflectance experiment, we consider a single slit in a gold layer deposited on a glass substrate. The slit is illuminated from the rear glass/gold interface. The incident light is dominantly reflected or absorbed. Only a small fraction of the incident energy is funneled through the slit. At the front aperture, light is either diffracted in the far field or dissipated as heat at the air-gold interface, see Fig. 1a. A Scanning Electron microscope image of the slit is shown in Fig. 1b. To measure the absorbed power on the top interface, any undesirable background side effects induced by diffusion of the large amount of heat deposited by the pump on the rear side should be avoided. Bearing this in mind, we use femtosecond pulsed lasers and we design a specific multilayered system.

A first laser pump pulse ($\lambda$ = 800 nm, 180-fs full-width-at-half-maximum) is slightly focused on the rear interface. The beam waist is 40 µm, a value slightly larger than the slit length. The polarization can be set either parallel (TE) or perpendicular (TM) to the slit. A second laser pulse ($\lambda$ = 532 nm, 180-fs FWHM, 2.6-µm beam-waist) probes the air-gold interface and records the reflectance variations $\Delta R$ with a controlled delay between the pump and the probe pulses.

Absorption of laser pulses by metal films can be described as a three-step process with different time scales [8-9]. Figure 1c shows a typical temporal $\Delta R$ response recorded on a 200-nm-thick gold mono-layer when the pump and probe beams spatially overlap. First, the energy is absorbed by the free electron gas. The latter is much lighter and reactive than the ion lattice and can thermalize very rapidly towards the equilibrium Fermi-Dirac distribution with a time scale of $\approx$ 100 fs. Then, the hot electron gas relaxes through internal electron-phonon collisions characterized by a time scale $\approx$ 1ps, before a classic heat diffusion transport takes place with a much longer characteristic time. For accuracy purposes, we decided to probe the electron-gas heating at the maximum of the $\Delta R$ signal in Fig. 1c. Thus all the following measurements are performed at a fixed pump-probe delay of $\approx$ 500 fs. Details on the subpicosecond pump-probe thermoreflectance setup can be found in [10].

Slits with different widths from $w$ = 200 to 800 nm are etched through a multilayered Au(150 nm)/SiNx (25 nm)/Au(100 nm)/Ti(5nm) film deposited on a glass substrate, see the inset in Fig. 1a. First, the Ti-adhesion and Au layers are sputtered on a fused-silica substrate surface cleaned with Ar-magnetron plasma sputtering. Afterwards, a SiNx layer is deposited with a Plasma-Enhance-Chemical-Vapor-Deposition technique, onto which the upper gold layer is sputtered. The slit is etched into the multilayer film with an electron beam lithography and argon-ion beam etching in an oxygen plasma. A 300-nm-width slit, which supports a single propagative $TM_0$ mode at the pump frequency, is used for the experimental results reported hereafter, but similar results were obtained with other sub-wavelength slit widths. The SiNx layer buried between two optically-thick gold layers is essential for the thermoreflectance measurements. This insulating layer acts as a barrier, to prevent the tremendous amount of heat deposited by the pump on the rear side. It blocks the electron and phonon flow coming from the rear interface, at least in the measured time window (500fs) of the pump-probe experiment. Therefore this allows us to reveal the weak absorption due to the electromagnetic field scattered by the slit at the front-side interface with thermoreflectance.

Figure 2a shows the thermoreflectance images of the front interface for TM and TE polarizations. For both polarizations, the thermoreflectance signal is maximal at the slit, where the pump energy is funneled and strongly heats the slit walls. It is also striking that, away from the slit, the reflectance variation is much weaker for TE polarization than for TM, consistently with the fact that, only for TM polarization, part of the scattered light is launched on the metal surface and is dissipated as heat. We believe that Fig. 2a reports the first "direct" nonintrusive observations of absorption loss induced by light scattering at a sub-wavelength aperture in metal films.

In order to obtain a more quantitative analysis, we extract 40 horizontal adjacent line scans from the central part of the TM image and average them. The averaged thermal profile is shown in Fig. 2b. As before, the absorption is maximal at the slit location. Far from the slit, the absorption gradually decreases. This behavior is expected since it is generally admitted that the amplitude of waves generated by local scatterers decreases as the distance increases. However, at intermediate distances from the slit, surprisingly, the absorption exhibited a remarkable feature in form of a plateau. This plateau reveals a nearly constant absorption over a large spatial interval (15 µm ≈ 18 λ). This implies that as it propagates away from the slit, the electromagnetic field launched at the metal interface heats the metal but remains essentially constant. Indeed, one would rather expect that this field decreases (due to source-distance increase and absorption). This paradoxical behavior will be explained hereafter.

The observed plateau depends on the spatial shape of the heat source (the absorbed electromagnetic wave), but it also depends on the complex heat transport that takes place in the metal during the experiment temporal-window. In order to show that the first effect is largely dominant, we carry out a numerical simulation, using a fully-vectorial aperiodic-Fourier-modal method [11] to compute the heat source profile. For the values of the gold permittivity $\varepsilon_m$ tabulated in [12], we calculate the electric field $\mathbf{E}(x,z)$ scattered by a slit aperture illuminated by the fundamental $TM_0$ slit mode – indeed there is no need to model the whole multilayered structure. Figure 3a shows the absorbed power density $A(x,z) = \frac{1}{2}\omega\, Im(\varepsilon_m)\|\mathbf{E}(x,z)\|^2$ in the metal for the parameters used in the experiment, $w = 300$ nm and $\lambda = 0.8$ µm. We further compute the vertically-integrated absorption profile $A(x) = \int A(x,z)dz$, where the integration is performed over a metal thickness of 110 nm (≈ 10 skin depths at $\lambda = 0.8$ µm). Consistently with the experimental observation, the profile A(x) displays a plateau that is almost constant over a 10-µm spatial interval. Note that A(x) has been convoluted by a Gaussian with a FWHM of 2.8 µm to mimic the impact of the finite probe-beam waist on the thermoreflectance measurements. The convoluted data are further fitted with the experimental data and are shown with the solid red curve in Fig. 2c. Conclusively, quantitative agreement is achieved.

We have performed the same calculations for other wavelengths, see Fig. 3b. The plateau size is found to increase with the wavelength, to reach a size of 70 λ at λ = 1.5 µm. In the visible for λ ≤ 0.7 µm, A(x) rapidly decreases with x and no plateau is observed. For λ ≥ 1.5 µm, most of the energy is dissipated in the immediate vicinity of the slit, and the plateau is still observed but for very weak absorptions.

From the computed scattered field $\mathbf{E}(x,y)$, we further extract the electric field $\mathbf{E}_{SPP}(x,y)$ of the SPPs that are launched on each side of the slit, to further compute the SPP-absorption profile $A_{SPP}(x) = \frac{1}{2}\omega\int Im(\varepsilon_m)\|\mathbf{E}_{SPP}(x,z)\|^2 dz$ that would be absorbed if the electromagnetic field in the metal would only be composed of SPPs. Details on the method used can be found in [13]. $A_{SPP}(x)$ is shown with red-dashed curves in Fig. 3b and as expected, decreases exponentially with x. At long distances from the slit, A(x) ≈ $A_{SPP}(x)$, which implies that the absorption is mainly due to the launched SPPs. However in the plateau area, we note that the actual absorption is substantially smaller than that of the SPP, implying that *only a fraction of the launched SPP field is actually dissipated as heat.*

In order to gain physical insight into this anomalous behavior and to understand how the plateau varies with material parameters and wavelength, we adopt the dual-wave picture in [7] and assume that the electromagnetic field scattered by the slit is composed of SPPs and quasi-cylindrical waves. Accurate analytical expressions obtained by integration in the complex plane are known for the quasi-cylindrical waves [7], but for simplicity we use a very simple expression with a $x^{-3/2}$ damping with the distance. Thus, on the right side of the slit (x > 0) and in the metal (z < 0), the two-row electric-field vectors $[\mathbf{E}_x, \mathbf{E}_z]$ of each wave are written

$$\mathbf{E}_{SPP}(x,y) \prec exp(ikn_{SPP}\, x - ik\chi_{SPP} z)[\chi_{SPP}, n_{SPP}], \tag{1}$$

$$\mathbf{E}_{CW}(x,y) \prec exp(ikn_{CW}\, x - ik\chi_{CW} z)(kx)^{-3/2}[\chi_{CW}, n_{CW}], \tag{2}$$

for the SPP and quasi-cylindrical waves, respectively. In the previous equations, $k = \omega/c$ is the wave-vector modulus, $\varepsilon_d$ is the permittivity of the dielectric, $n_{SPP} = (1/\varepsilon_m + 1/\varepsilon_d)^{-1/2}$ and $n_{CW} = (\varepsilon_d)^{1/2}$ are the normalized propagation constants of the SPP and quasi-cylindrical waves, $\chi_{SPP} = (\varepsilon_m - (n_{SPP})^2)^{1/2}$, $\chi_{CW} = (\varepsilon_m - \varepsilon_d)^{1/2}$ and $\gamma = n_{SPP} - n_{CW}$. With the approximations $\chi_{SPP} \approx \chi_{CW}$ and $n_{SPP} \approx n_{CW}$, it is straightforward to show that the vertically-integrated absorption profile $A(x) = \frac{1}{2}\omega \int Im(\varepsilon_m) \|\mathbf{E}_{SPP}(x,z) + \mathbf{E}_{CW}(x,z)\|^2 dz$ can be written as

$$A(x) \approx A_{SPP}(x)\left[1 + 2\sqrt{\frac{A_{CW}(x)}{A_{SPP}(x)}}\cos(Re(kn_{SPP} - kn_{CW})x + \alpha)\right] + A_{CW}(x). \tag{3}$$

In Eq. (3), $A_{SPP}(x) = \frac{1}{2}\omega \int Im(\varepsilon_m)\|\mathbf{E}_{SPP}\|^2 dz = A_{SPP}(0)\exp(-2k\,Im(n_{SPP})x)$ is the exponentially-decaying absorption due to the SPP, $A_{CW}(x) = \frac{1}{2}\omega \int Im(\varepsilon_m)\|\mathbf{E}_{CW}\|^2 dz = A_{CW}(0)(kx)^{-3}$ is the algebraically-decaying absorption due to the quasi-cylindrical wave, which can be neglected a few wavelengths away from the slit. $A_{SPP}(0)$, $A_{CW}(0)$ and $\alpha$ are three real independent parameters that are fitted, using the numerical data obtained with the aperiodic-Fourier-modal method. The red dots in Fig. 3b are fitted values computed from Eq. (3). For all wavelengths, they well match the numerical data. Despite its simplicity, the analytical model well predicts all the salient features of the absorption profile, the initial rapid decrease followed by a plateau, and the exponential decrease at large separation distances.

The "anomalous" absorption profile is thus understood as resulting from a background term, the SPP absorption $A_{SPP}(x)$, over which an oscillatory term is imprinted with a spatial frequency that results from the beating of a SPP and a quasi-cylindrical wave. Noting that the plateau length $L$ is approximately given by half the beating period, we obtain a very simple expression for $L$

$$\frac{L}{\lambda} = \frac{\pi}{2\,Re(n_{SPP} - n_{CW})} \approx Re\left(\frac{\varepsilon_m}{\varepsilon_d^{3/2}}\right). \tag{4}$$

We have verified that Eq. (4) represents a very good approximation of the plateau size for 0.8 μm < $\lambda$ < 1.5 μm and for various dielectrics ($\varepsilon_d$ = 1, 2.25, 4), and noble metals (gold, silver).

For the sake of generalization, point-source (instead of line-source) scatterers are now investigated. Instead of a specific geometry such as a nano-hole in a metal film, we rather consider electric-dipole sources located just above a flat gold surface and calculate the absorbed power-density profile $A(\rho,\theta) = \frac{1}{2}\omega \int Im(\varepsilon_m)\|\mathbf{E}(\rho,\theta,z)\|^2 dz$ radiated by the sources in the metal ($\rho$ and $\theta$ denotes the radius and the azimuth in cylindrical coordinates). $A(\rho,\theta)$ is shown in Fig. 4a for two dipole sources emitting at $\lambda$ = 0.8 μm, respectively polarized perpendicularly and parallel to the surface along the x-axis. In Fig. 4b, the solid curves represent $\rho A(\rho,\theta)$, a quantity directly proportional to the absorbance of the surface element $\rho d\rho d\theta$ located at polar coordinate $(\rho,\theta)$, as a function of $\rho$ for a dipole parallel to the surface and for four values of $\theta$, $\theta = \pi/6, \pi/4, 2\pi/5$ and $\pi/2$. Remarkably, all curves exhibit a large plateau. The same conclusion holds for the azimuthally-averaged absorbance $A(\rho) = \int A(\rho,\theta)\,d\theta/2\pi$ of an elementary ring of radius $\rho$ (see the circles in the Fig. 4b). Virtually identical results (upon rescaling) are obtained for the perpendicular polarization. We finally note that the absorbance profiles $A(\rho)$ in Fig. 4b and $A(x)$ in Fig. 3d, which are obtained for point and line sources respectively, are virtually superimposed when rescaled and plotted on the same graph at $\lambda$ = 0.6, 0.8, 1 and 1.5 μm (see the dotted-blue curve and the associated circles in Fig. 4b). This leads us to the conclusion that *the anomalous absorption observed experimentally for slits is a general property of subwavelength apertures in metal films.*

In summary, we have provided the first comprehensive study of the photothermal heat generation in plasmonic films incorporating subwavelength indentations. We have shown that the field scattered by point or line scatterers on the surface of metal films remains constant over a large scale of several tens of wavelengths. This phenomenon that occurs with noble metals at near-infrared wavelengths (λ > 800 nm) is an intrinsic property, which is likely to be observed for any subwavelength aperture geometries. In the plateau area, the dissipated heat is weaker than the heat deposited by the plasmon launched at the surface. This property opens interesting perspectives for optimizing the performance of metallic devices for energy harnessing in complex plasmonic systems that combine surface plasmons and localized plasmonic resonances [14]. Alternatively it may be used to derive new recipes for designing compact plasmon launchers that deliver surface plasmons with a reduced heat dissipation at the delivery point, a valuable property that may be used in quantum plasmon circuitry for lowering decoherence [15].

This study was carried out with financial support from "the Investments for the future" Programme IdEx Bordeaux – LAPHIA (ANR-10-IDEX-03-02).

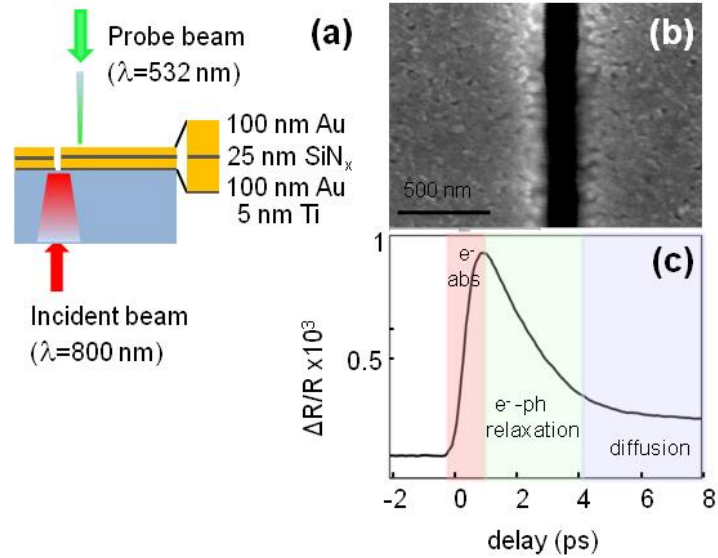

**FIG. 1 (a)** Sketch of the sub-picosecond pump-probe slit-experiment. The inset shows the multilayer structure of the sample. **(b)** Scanning Electron microscope image of the slit used in the experiment. **(c)** Time-resolved response of a 200-nm-thick gold layer as a function of the pump-probe delay.

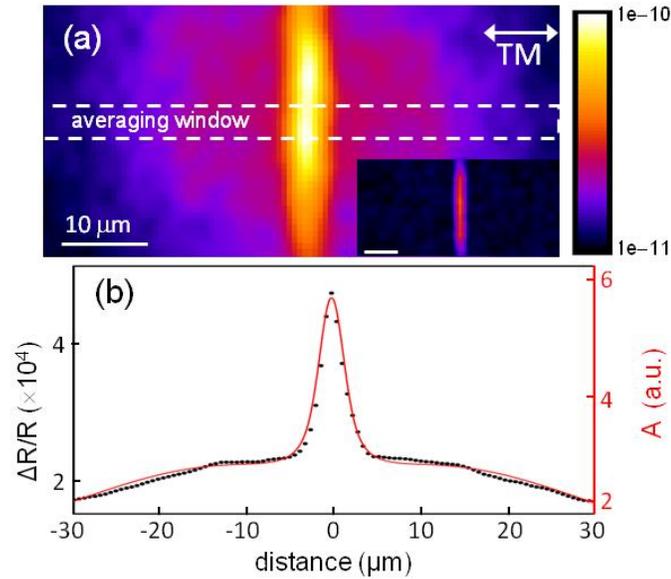

**FIG 2.** Thermoreflectance data for a 300-nm-width slit. (**a**) Images obtained for TM polarization (the inset is obtained for TE). **(b)** Data (black dots) are obtained by averaging 40 line scans from **(b)**. The solid red curve represents the calculated absorption profile $A(x)$ convoluted by a Gaussian with a FWHM equal to the waist (2.8 µm) of the probe beam.

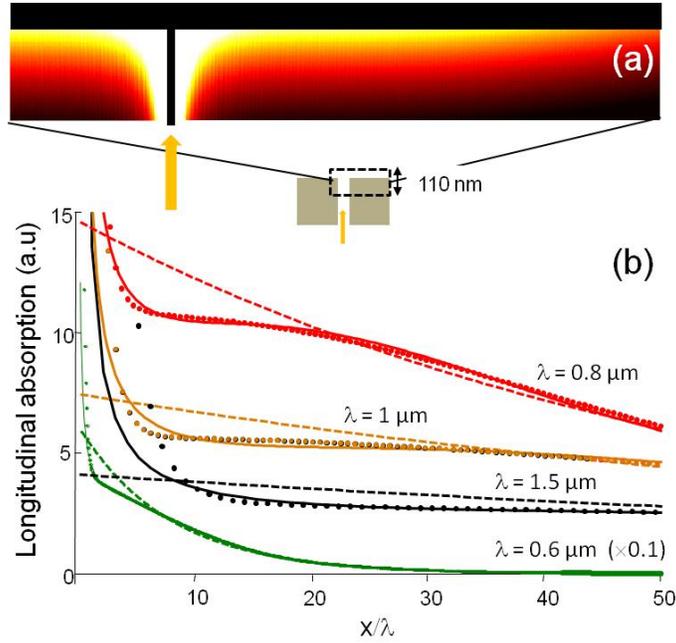

**FIG 3.** Computed results obtained for the absorption at a gold-air interface illuminated by the fundamental $TM_0$ mode of a slit. **(a)** Absorbed power density $A(x,z)$ in a linear scale at $\lambda$ = 800 nm for $w$ = 300 nm. **(b)** The solid curves represent the absorption profile $A(x) = \int A(x,z)\, dz$. The dashed curves represent the absorption profile $A_{SPP}(x)$, which would be achieved if the metal field was only composed of two SPPs launched in opposite directions. Results are presented for several wavelengths and for the same slit-width-to-wavelength ratios, $w/\lambda$ = 3/8. The dots are fitted data, using Eq. (3).

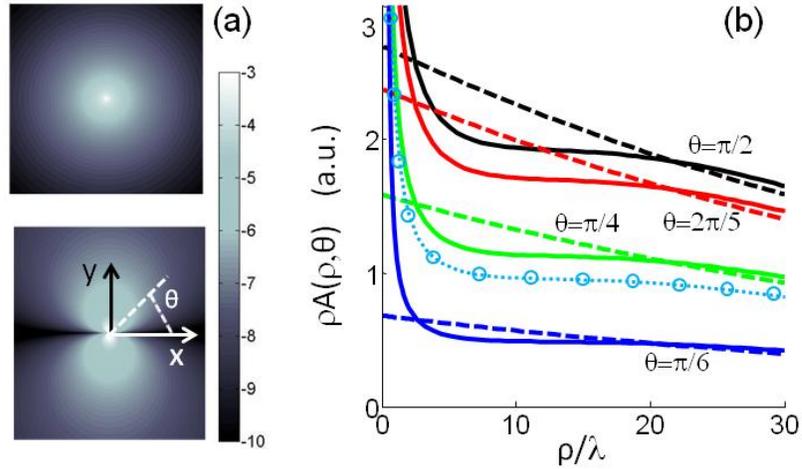

**FIG 4.** Absorbance at a gold-air interface illuminated by a point dipole source located at x=y=0, just above (z=0$^+$) the metal plane and emitting at $\lambda$ = 800 nm. **(a)** Absorbance $A(\rho,\theta)$ for a dipole perpendicular (upper panel) or parallel (lower panel) to the surface and oriented along the x axis. **(b)** Solid curves: $\rho A(\rho,\theta)$ for several $\theta$. The dashed curves correspond to the associated SPP absorbance $\rho A_{SPP}(\rho,\theta)$. The dotted-light-blue curve represents the azimuthally-averaged absorbance $A(\rho)$. It is virtually superimposed with the circles that represent the absorption profile $A(x)$ of the slit (red curve in Fig. 3b) after vertical rescaling.